\documentclass[aps,showpacs,preprint,preprintnumbers,amsmath,amssymb]{revtex4-1}

\usepackage{amssymb}
\usepackage{amsmath}
\usepackage{graphicx}
  \usepackage{url}
\usepackage{hyperref}

\begin{document}
\title{Reply to
 "Comment on: Magnetotransport through graphene spin valves and its following works" \\ by Y. Zhou and M.W. Wu}

\author{K. H. Ding$^1$ Z. G. Zhu$^2$, and J. Berakdar$^2$}
\affiliation{
$^1$Department of Physics and Electronic Science, Changsha University of Science and Technology, Changsha,410076, China\\
$^2$Institut f\"{u}r Physik, Martin-Luther-Universit\"{a}t Halle-Wittenberg,  06099 Halle (Saale), Germany
}

\begin{abstract} In their comment
 Y. Zhou  and M.W. Wu claim that the fundamental  transport equation relating the current to the transmission function,
used by us and in fact by numerous other researchers, is invalid  for extended systems  and should be corrected.
 They provide a "correct" new formula for transport in extended systems.
  This would be indeed a surprising new aspect of
 quantum transport theory. Here we show mathematically, however,  that
the "new formula" is a   misconception resulting from adding an energy and momentum dependent function
that has to vanish, due to fundamental reasons. Results and conclusions stemming from adding this function are
irrelevant. The known
established formulas for quantum transport are consistent with each others under  the well-documented conditions.
\end{abstract}

\maketitle

In their comment \cite{ZW} %
 Y. Zhou  and M.W. Wu argue that   the steady-state current $I_L$ in an extended quantum system
connected to two leads $L$ and $R$ evaluated according to \cite{fisher,baranger,caroli,imry,dattabook,haug}
\begin{equation}
I_L=I_{LB}= \frac{2 e}{h} \int_{-\infty}^\infty d\epsilon\;  tr \left\{(f_L- f_R) \left( \Gamma_L G^+\Gamma_R G^-\right)\right\}
\label{eq1}
\end{equation}
is  not consistent with $I$ when calculated with the other established formula \cite{MW,dattabook,haug}
\begin{equation}
I_L=I_{MW}= \frac{ie}{h} \int_{-\infty}^\infty d\epsilon\; tr \left[
(f_L\Gamma_L - f_R\Gamma_R) (G^+-G^-) +   (\Gamma_L - \Gamma_R) G^< \right]
\label{eq2}
\end{equation}
(for finite systems they claim $I_{LB}=I_{MW}$).
We use the standard notations where  $f_\alpha\equiv f_\alpha(\epsilon -\mu_\alpha)$ is the distribution function
on the left ($\alpha=L$) or  right ($\alpha=R$) lead with the chemical potential $\mu_L$ ($\mu_R$) and $\Gamma_\alpha= i\left( \Sigma^+_\alpha -
\Sigma^-_\alpha\right)$, where $\Sigma_\alpha^{\pm}$ is the respective  selfenergy. The superscript $+$ ($-$)
stands for retarded (advanced) quantities. The operators $G^{\pm}$ satisfy a Dyson equation that can be written as
\begin{equation}
G^{\pm}=\left[   {g_c^\pm}^{-1}
-(\Sigma_L^\pm + \Sigma_R^\pm) \right]^{-1}
\label{eqG}
\end{equation}
 hereby $g_c(z)=(z-H_c)^{-1}$ is the resolvent of
 $H_c$, which is a hermitian Hamiltonian describing
  the isolated central region and $g_c^\pm(z)= g_c(\epsilon \pm i\eta )$ where $\eta$ is a small positive real number taken to zero \emph{after}
  performing the trace and the energy integration in eq.(\ref{eq1},\ref{eq2}).
$G^<= i G^+\left( f_L \Gamma_L +f_R \Gamma_R\right) G^-$ (cf. \onlinecite{haug}).  \\
Zhou and Wu claim that our previous results \cite{ding2009,dingall} calculated with eq.(\ref{eq1}), and for that matter
the results of all other researchers employing the same approach for an extended system, lack scientific ground.
Let us show mathematically that the inconsistencies found by Zhou and Wu  when using eq.(\ref{eq1}) vs.\ eq.(\ref{eq2})
are  self-made  and indeed eq.(\ref{eq1}) and eq.(\ref{eq2})
should yield consistent results independent of whether the system is finite or extended (in the sense introduced by Zhou and Wu in Ref.[\onlinecite{ZW}]).

For clarity let us work in a representation free manner and write the operator equation
 $(G^+-G^-)=G^+\left({G^-}^{-1}-{G^+}^{-1}\right)G^-$ which readily yields (cf. eq.[\ref{eqG}])
\begin{eqnarray}
G^+-G^-&=&G^+\left[ (\epsilon -i\eta - H_c) - (\epsilon +i\eta - H_c)\right] G^-  - i G^+(\Gamma_L+ \Gamma_R)G^- \label{eq3_0}\\
&=& -2i\eta G^+G^- - i G^+(\Gamma_L+ \Gamma_R)G^- .
\label{eq3}\end{eqnarray}

The trace of this equation is to be compared  with the eq.(5) in Ref.[\onlinecite{ZW}].
 Zhou and Wu claimed  \emph{that the first term in Eq. (\ref{eq3_0}) vanishes
... in the finite system, in consistence
with the previous literature. However, in the infinite
system discussed by Ding {et al.} \cite{ding2009} the situation becomes
totally different}.  Hence, they derive a "new correct formula" for extended systems
 by taking the first term of Eq. (\ref{eq3}) into account
and construct a way to make it  finite.

As a matter of fact for a finite $\Gamma_\alpha$, mathematically
 the trace  of the first term
  of Eq. (\ref{eq3})
  \begin{equation}
-2i\eta G^+G^- = G^+\left[ (\epsilon -i\eta - H_c) - (\epsilon +i\eta - H_c)\right] G^-
\label{equation31}\end{equation}
   has  to vanish always and in any basis  when $\eta \to 0$, for
in this case $G^\pm$ has neither isolated poles nor a branch cut for $\eta \to 0$, i.e. when approaching the real energy axis.
 This is also evident from the  structure of $G^\pm$ (cf. \ref{eqG}).
For $\Gamma_\alpha\to 0$, the trace over $-2i\eta G^+G^-$ yields  for $\eta\to 0$  indeed the spectral density of the system,
and the second term of Eq. (\ref{eq3}) is identically zero. This
is consistent with the well-established meaning of the trace over $G^+-G^-$.
 For $\Gamma_\alpha\to 0$ however
 the current $I_{MW}$ vanishes as clear from eq.(\ref{eq2}). This means in turn that introducing somehow a
 finite trace over $-2i\eta G^+G^-$ for $\eta\to 0$ regardless of  $\Gamma_\alpha$ being finite, amounts to a change of the
 system spectral density and raises thus the question of the charge conservation (i.e. $I_R=-I_L$).
  Indeed, as well-established and readily
 deducible from both
 prescriptions (\ref{eq1}) and (\ref{eq2})  the charge conservation is fulfilled for (\ref{eq1}) and (\ref{eq2}).
 Constructing somehow  a finite trace of the term (\ref{equation31}) one may enforce as an additional condition that
  $I_R+I_L=0$ and distribute accordingly the spurious term on $I_L$ and $I_R$, 
  such
an approach to restore the charge conservation, however,
 is far from
  being fundamental!.

We infer mathematically thus that
 for steady-state transport, i.e. when $\Gamma_\alpha$ is finite, the first term of  Eq. (\ref{eq3_0})
 plays no role.

Nonetheless,  Zhou and Wu argue that the term $-2i\eta G^+G^-$ should be finite because in eq.(\ref{eq3_0}) one may write
(cf. eq.(5) in Zhou and Wu comment)
\begin{equation}
G^+\left[ (\epsilon -i\eta - H_c) - (\epsilon +i\eta - H_c)\right] G^-=G^+\left[ {g_c^-}^{-1} - {g_c^+}^{-1}\right] G^-
\label{eqzw}\end{equation}
and assume a finite   (${g_c^-}^{-1} - {g_c^+}^{-1}$).
 While one may do such a manipulation the contribution of this term  to the current
remains of course zero \cite{footnote0}.
Clearly, replacing  a vanishing term by an energy and momentum dependent function
may lead to a series  of conclusions that are at variance with known results,
including   the statement that eq.(\ref{eq1}) is not applicable for an extended system.

   The matter of fact however, for a finite $\Gamma_\alpha$  only the second term of eq.(\ref{eq3}) contributes when taking the trace
   in eqs.(\ref{eq1},\ref{eq2}) and  letting $\eta\to 0$. It is straightforward to show   by inserting
   eq.(\ref{eq3}) into eq.(\ref{eq2}) that one retrieves   the established result
\begin{equation}
I_L= \frac{e}{h} \int_{-\infty}^\infty d\epsilon\; tr \left\{
(f_L- f_R) \left[\Gamma_LG^+\Gamma_R G^- + \Gamma_R G^+\Gamma_L G^-  \right]\right\}
\label{eq4}
\end{equation}
which is equivalent to  eq.(\ref{eq1}) that we and others use for
   the calculation of the steady-state current.

 Hence, as far as the system size is concerned, as introduced by Zhou and Wu and we only focus on this issue here,
  one may use  eq.(\ref{eq1}) or eq.(\ref{eq2}) and finds $I_{LB}=I_{MW}$. More importantly any
 effects on the transport  based on a finite first term in eq.(\ref{eq3}) should be considered artificial and
 resulting  from  some uncontrolled approximations.   Based on their "correct" formula Zhou and Wu
 go even a step further in their conclusions and state
 that works for an extended system employing eq.(\ref{eq1}) are incorrect and lack scientific ground.
 This statement is clearly a consequence of a self-made finite term that for a finite current should be in fact zero due to
 fundamental reasons.

\begin{figure}[tbh]
\includegraphics[width=0.8\textwidth]{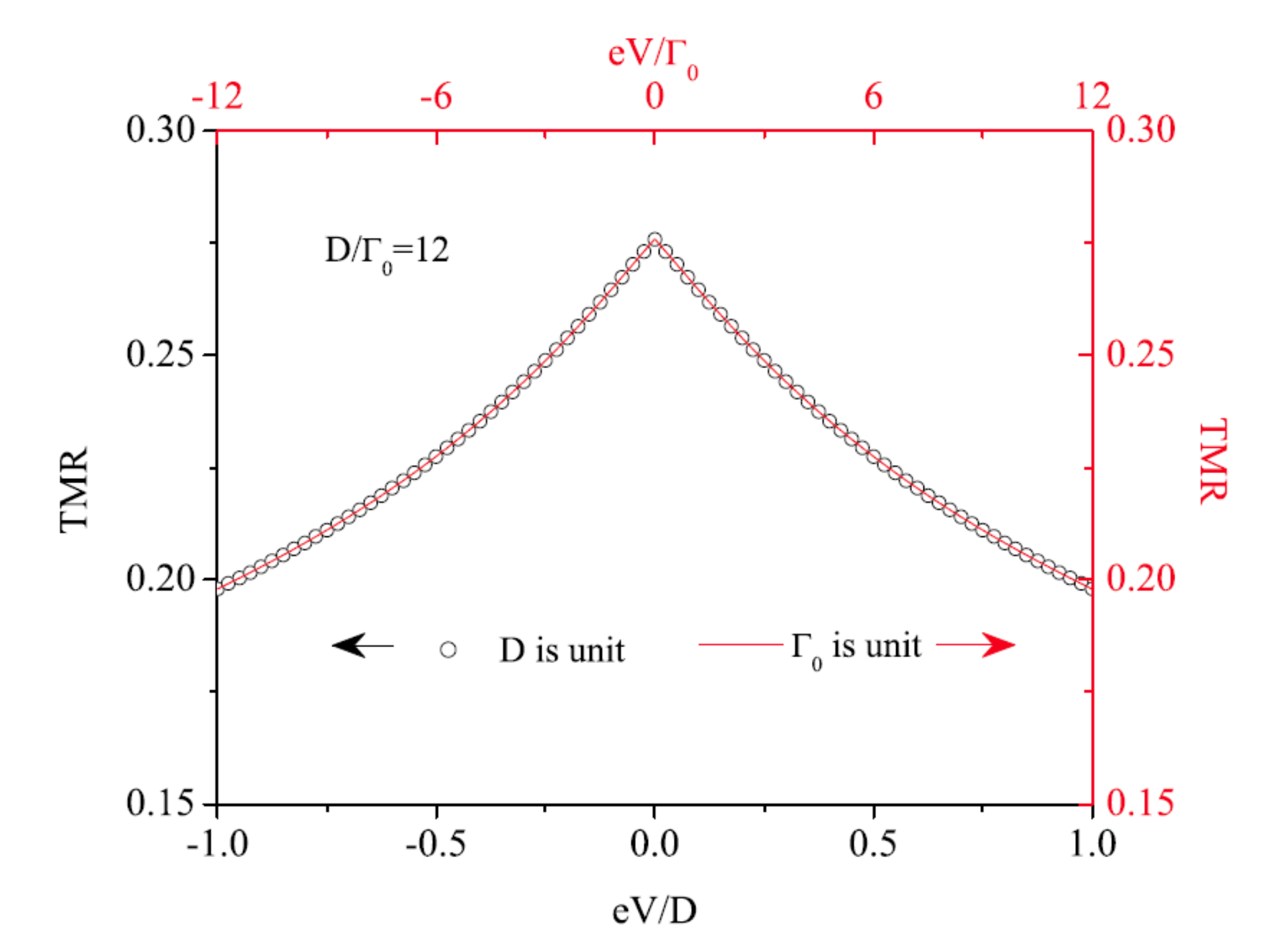}
\caption{(color online) TMR in a graphene monolayer in contact with two metallic ferromagnetic leads
as a function of the applied bias V, calculated according to eq.(\ref{eq1}) or eq.(\ref{eq2})  while correctly
neglecting the first term of eq.(\ref{eq3}) (as done in  our previous work \cite{ding2009}).
 The dots are the results derived by setting the energy width $D$ as the unit of energy. The solid line is the result by using the broadening
 function $\Gamma_{0}$ as the unit of energy. $D/\Gamma_{0}=12$ and the spin polarization of the two metallic ferromagnetic leads is
  $40\%$. The leads are assumed to be of the same material.
\label{TMR}}
\end{figure}
\begin{figure}[tbh]
\includegraphics[width=0.8\textwidth]{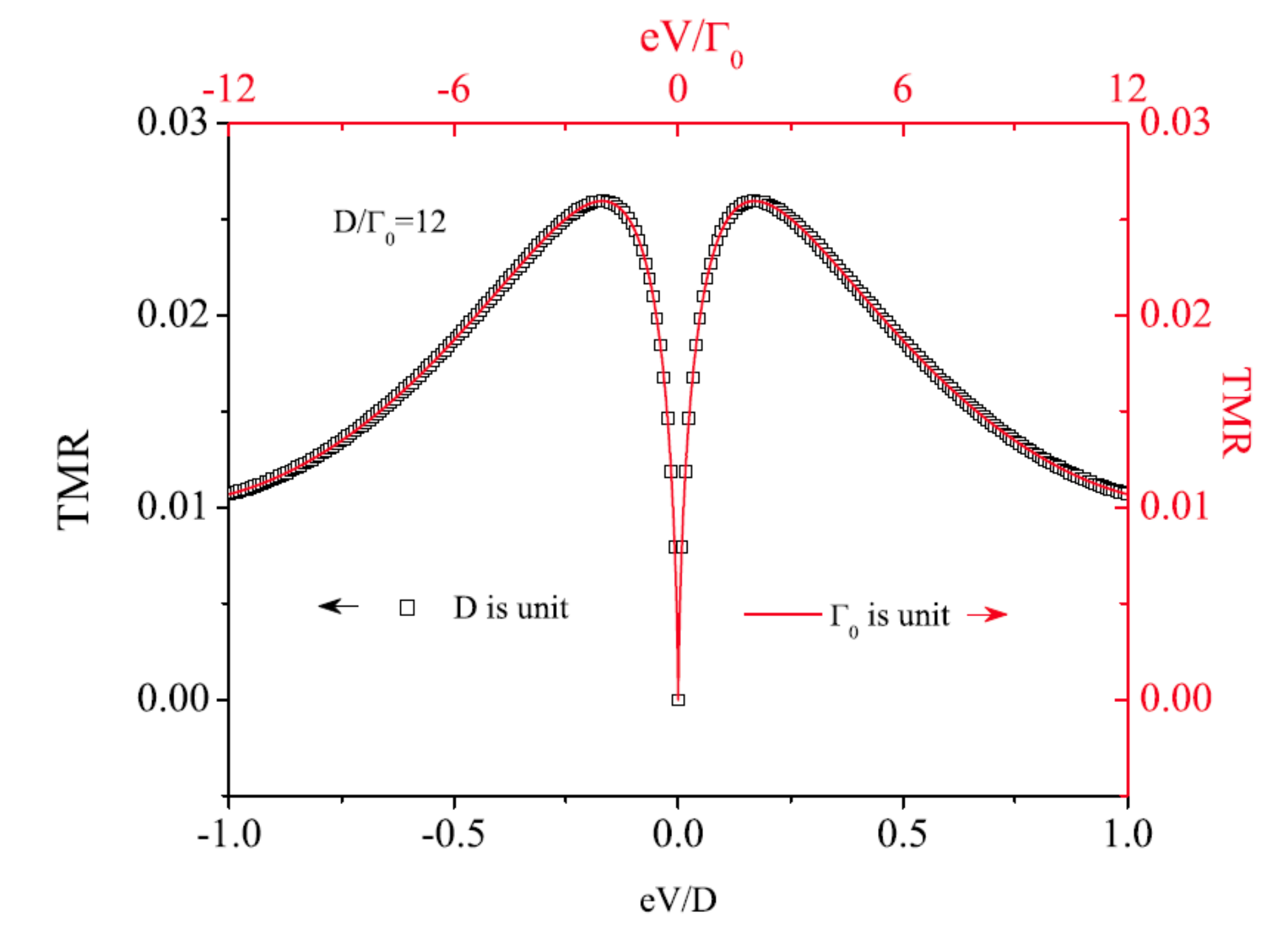}
\caption{(color online) as in Fig. \ref{TMR}, however the TMR is calculated by using the suggestion by Zhou and Wu in Ref.[\onlinecite{ZW}].
 The parameters are the same as those in Fig. \ref{TMR}.
\label{TMR1}}
\end{figure}

  Based on their new, allegedly "correct" formula Zhou and Wu raise some issues concerning our results \cite{ding2009,dingall}, in particular
  those for  the tunnel magnetoresistance (TMR)
  of a graphene monolayer contacted to metallic ferromagnetic leads that we obtained
  on the basis of eq.(\ref{eq1}). Since our results do not agree with their calculations based on their own constructed formula
  they claim  that the reason is  due to a "wrong" energy cutoff  $D$ that we use in our calculation. It should be noted, that
  as shown by Zhou and Wu the spurious term they include in their formula contains $D$ in a non-trivial manner.
  In their comment \cite{ZW} they make several claims based on the dependence of their calculated current on $D$.
   While this point is somehow technical, a clarification might be useful to avoid
  a misunderstanding of the meaning of $D$.
  We choose $D$
as to ensure the
conservation of the
number of states in the Brillouin zone (upon linearizing  the spectrum, cf. Ref.[\onlinecite{uchoa2008}]).
Calculating $D$ accordingly one arrives at the Green function given in our works. This is a physically motivated way to choose $D$
that can be set as the energy scale. Of course, one may choose another $D$ which in turn means a violation of the
number of states in the Brillouin zone.

 As shown in Figs.\ref{TMR},\ref{TMR1},  we can perform the calculations using $D$ or $\Gamma_0$
 ($\Gamma_\alpha$ are assumed to be momentum and energy independent)
 as the energy scale and arrives at the same behaviour of the TMR.
 Proceeding as suggested in the comment 
  by  Y. Zhou  and M.W. Wu  one arrives at an opposite  physical behaviour
 of TMR, i.e.
 a zero TMR at zero bias instead of a peak. Indeed, it is straightforward to show analytically, that this behaviour is a direct consequence
 of assuming, as done by Zhou and Wu,  a finite first term in eq.(\ref{eq3_0}) that is related to   the spectral density of graphene.
 In addition, in their comment Zhou and Wu show the result of their "correct" formula for the conductance.
 It  can be shown mathematically that the behaviour of the conductance  at small bias in their case is dominated
 by the erroneous  finite first term in eq.(\ref{eq3}).

In Summary, eq.(\ref{eq1}) and eq.(\ref{eq2}) are valid irrespective of whether the system is finite  or extended (in the sense metioned in the comment).
The claims of  Y. Zhou  and M.W. Wu in their comment
are the result of a fabricated finite energy and
momentum-dependent term that should vanish
due to fundamental reasons if the current is finite.
Established approaches to quantum transport are consistent within the well-documented  limits.\\

Note, we do not consider Eq. (18) of Zhou and Wu in our numerics. We stress that, as stated in
the figure caption, our Fig. 2 is obtained as Fig. 1 from our
theory, but we include in the calculations a constructed finite
spurious term of the form given by Eq. (7), along the line as we
understand the suggestion by Zhou and Wu. Note, as discussed above,
 in this case the charge conservation is not a priori guaranteed.


\begin{thebibliography}{}
\bibitem{footnote0}
This is also obvious when working with the momentum ($\mathbf q$) states of graphene, in which  case
the  graphene Green's is diagonal and reads
 \begin{equation}
g^{\pm}_{\mathbf{q},\mathbf{q}}(\varepsilon)=\frac{1}{\varepsilon\pm i\eta-q_x\sigma_x-q_y\sigma_y}
\end{equation}
and the Green function for the composite structure satisfies the Dyson equation
\begin{equation}
G^\pm_{\mathbf{q},\mathbf{q}'}(\varepsilon)=\delta_{\mathbf{q}\mathbf{q}'}g^\pm_{\mathbf{q},\mathbf{q}}(\varepsilon)
+\sum\limits_{\mathbf{q}_1}g^r_{\mathbf{q},\mathbf{q}}(\varepsilon)\sigma^+\Sigma_{\mathbf{q}\mathbf{q}_1}^\pm(\varepsilon)
G^\pm_{\mathbf{q}_1,\mathbf{q}'}(\varepsilon)\label{grq}
\end{equation}
where $\sigma^+=\frac{1}{2}(1+\sigma_z)$.
With these expressions, it is readily inferred that  taking the trace in  (\ref{eqzw}) and letting $\eta \to0$ the  term
given by eq. (\ref{eqzw}) vanishes.
\end{thebibliography}
\end{document}